# Cultural context shapes the carbon footprints of recipes

Mansi Goel[1,2], Vishva Nathavani[1,3], Smit Dharaiya[1,3], Vidhya Kothadia[1,3], Saloni Srivastava[1,3] and Ganesh Bagler*[1,2]

[1]Infosys Centre for Artificial Intelligence
[2]Department of Computational Biology
[3]Department of Computer Science
Indraprastha Institute of Information Technology Delhi (IIIT-Delhi), New Delhi, India 110020.

*Corresponding author: Ganesh Bagler (bagler@iiitd.ac.in)

**Food systems are responsible for a third of global anthropogenic greenhouse gas emissions[1] central to global warming and climate change[2,3]. Increasing awareness of the environmental impact of food-centric emissions has led to the carbon footprint quantification of food products[4]. However, food consumption is dictated by traditional dishes[5], the cultural capsules that encode traditional protocols for culinary preparations. Carbon footprint estimation of recipes will provide actionable insights into the environmental sustainability of culturally influenced patterns in recipe compositions. By integrating the carbon footprint data of food products with a gold-standard repository of recipe compositions[6], we show that the ingredient constitution dictates the carbon load of recipes. Beyond the prevalent focus on individual food products[4], our analysis quantifies the carbon footprint of recipes within the cultural contexts that shape culinary protocols. While emphasizing the widely understood harms of animal-sourced ingredients[7,8], this article presents a nuanced perspective on the environmental impact of culturally influenced dietary practices[9]. Along with the grasp of taste[10] and nutrition[6] correlates, such an understanding can help design[11,12] palatable and environmentally sustainable recipes. Systematic compilation of fine-grained carbon footprint data[13] is the way forward to address the challenge of sustainably feeding[14] an anticipated population of 10 billion[15].**

Food systems contribute significantly to greenhouse gases (GHGs)[1] responsible for global warming and climate change[3,16,17]. Expressed in $CO_2$-equivalent, carbon footprint[18] (CF) captures the total amount of GHGs generated over the life cycle of an entity, primarily accounting for carbon dioxide, methane, and nitrous oxide emissions. It is directly linked to climate change with severe environmental and existential consequences for the planet. Understanding and mitigating the adverse environmental impact of global warming and climate change begins with quantification of carbon footprint. Estimating the carbon footprints of dietary consumption will provide better food choices toward a sustainable future.

Dominated by methane and nitrous oxide, food-centric GHG emissions[2,19] have origins in various factors from the food system lifecycle encompassing stages from farm to fork[20]. Besides deforestation, storage, transportation[21,22], and livestock activities[23], the energy needed to convert raw ingredients into edible food significantly contributes to GHGs. Global dietary consumption is dominated by food cooked using traditional recipe protocols.

Today, human civilization faces the challenge of making food sustainably accessible for an estimated ten billion world population anticipated by 2050[15]. Towards addressing this challenge, computational approaches[11] rooted in data of recipes[6] and the grammar of cooking[12] present an excellent opportunity for designing environmentally sustainable recipes while minimizing their carbon load. This combinatorial optimization problem requires blending ingredients to create tasty and nutritional dishes. A paramount question, therefore, is, 'How to combine ingredients to create recipes that are simultaneously palatable and environmentally sustainable?' Computational gastronomy[24] enables answering such questions of culinary origin through a data-driven approach by mining the underlying patterns in food, nutrition, cooking, and carbon footprint.

In this study, we aimed to compile the carbon footprint data of food ingredients and estimate the footprints of recipes from across global cuisines (Figure 1a). Creating a data repository of the carbon footprints of food ingredients and globally consumed recipes will enable making climate-friendly culinary choices. Such a resource will also enable the application of data-driven approaches for designing environment-friendly recipes and help facilitate sustainable food practices to minimize climate impact. Toward achieving this objective, we integrated carbon footprint data of food products from SU-EATABLE LIFE[4] with RecipeDB[6], a gold-standard repository of the ingredient composition of recipes. Our analysis provides insights into the environmental impact of recipes by examining the role of cultural practices and dietary preferences.

**Carbon footprint data of food products**

SU-EATABLE LIFE[4] (henceforth SuEatable) is the most extensive dataset of the carbon footprints of food products. It provides CF of food items grouped under broad typologies across four food groups (agricultural processed, animal husbandry, crops, and fishing) from 18 world regions and 112 countries (See Supplementary Table 1). These data points compiled from over 889 research articles provide the ecological impact of food ingredients in $CO_2$ equivalent per kilogram. 'Agricultural processed' and 'crops' commodity groups account for 64.56% of the food products, with the rest equally represented by 'fishing' and 'animal husbandry' (Figure 1b). While there is much to be desired in the consistency, geospatial resolution, and coverage of food products across diverse cuisines, this dataset presents the best open resource for ingredient-level CF data. We carefully processed these data to address various issues hindering its integration with RecipeDB–typos, redundant entries, incorrect mapping of regions and countries, and inconsistencies in formatting and data structure. Post-cleansing, we were left with 302 SuEatable food products.

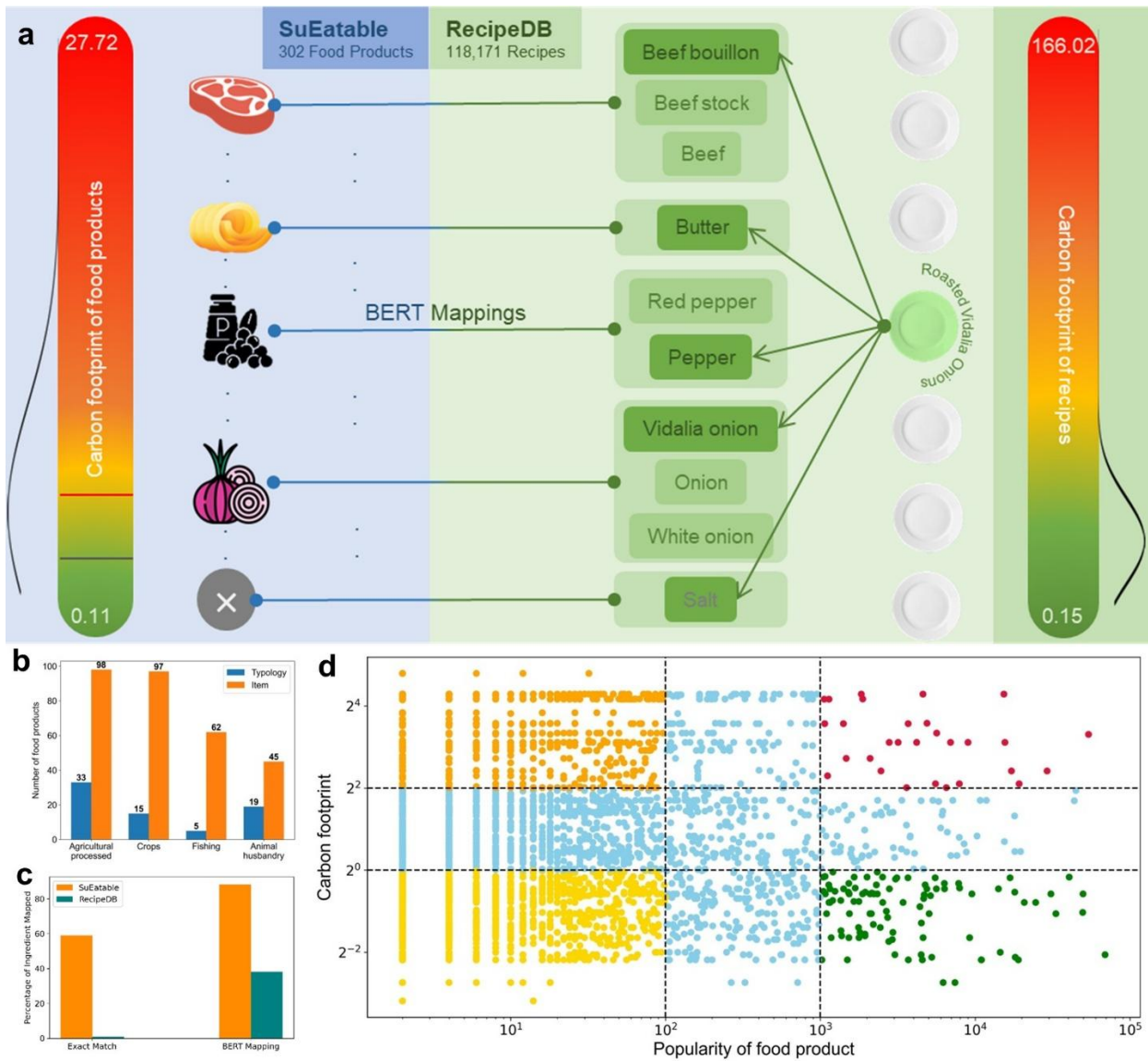

**Figure 1: Integrating carbon footprint data of individual food products (SuEatable) with recipe composition (RecipeDB) facilitates insights into the environmental impact of ingredients from their culture-driven popular use in culinary preparations.** (a) Schematic depicting the process involved in the integration of SuEatbale and RecipeDB data for the estimation of the carbon footprint of recipes. The lines adjacent to the color scale depict the nature of the corresponding data distribution. For the 'food products' the green and red lines depict CF<1 and CF>4 cut-offs, respectively. (b) SuEatable data statistics across food groups. (c) Ingredient mapping before and after BERT mapping, a computational protocol for meaningfully linking ingredients between SuEatable and RecipeDB. (d) Distribution of ingredients highlighting culturally popular ingredients with low (green) and high (red) footprints and rarely used ingredients that have low (yellow) and high (orange) ecological impact.

**Integration of SuEatable and RecipeDB**

To estimate the carbon footprint of a recipe, one needs to break it down into its constituent ingredients (food products) and map them to their respective footprints (Figure 1a). RecipeDB provides the ingredient constitution of 118,171 recipes from 26 regions[6]. For any recipe, mapping its RecipeDB ingredients with a relevant SuEatable 'food product' will help us determine its estimated carbon footprint. Therefore, we set out to map 20,280 ingredients from this structured recipe repository to the 302 food products (Figure 1a). The disparity in the number of entities between these databases is due to the many-to-one mapping of RecipeDB ingredients to that in SuEatable. Every SuEatable food product appears in different avatars in a recipe. For example, 'sugar' gets mapped to the following RecipeDB ingredients: white sugar, caster sugar, icing sugar, and confectioner sugar.

An exact match of the ingredient name with the food product label led to 178 entities getting mapped between RecipeDB and SuEatable (Figure 1c). Keeping cognizance of many-to-one mappings, we further implemented Word2Vec[25], RoBERTa[26], and BERT[27] language processing models to obtain the word embeddings for food product names in RecipeDB and SuEatable. We mapped the RecipeDB ingredient names to the SuEatable food item having maximum similarity between their embeddings. We compared the performance of models on the 200 most popular ingredients (with a similarity cutoff of > 81%). BERT was identified as the best model with accuracy and F1 scores of 81.5% and 86.32%, respectively. Figure E1 presents a detailed comparison of performance across the models. Figure E2 presents the success with which the most popular ingredients from RecipeDB were mapped to their correct analogs in SuEatable. Mapping with BERT-based word embeddings significantly improved the data integration with 7753 (38.2%) ingredient names from RecipeDB mapped to 266 (88.07%) SuEatable food items labels (Figure 1b). A significant proportion of popular ingredients, most frequently used in recipes, get mapped with their carbon footprints. See Supplementary Table 2 and Supplementary Figure 1 for a detailed comparison.

Among the most popularly used recipe ingredients successfully mapped after implementing the BERT-based mapping strategy were garlic clove, black pepper, lemon juice, vegetable oil, and parmesan cheese. This strategy also mapped many rarely used recipe ingredients such as walnut vinegar, welch grape jelly, ratafia biscuit, and gefilte fish. Despite the enhanced mappings post-implementation of the BERT-based mapping strategy, 36 SuEatable food items (coffee ground, haddock, krill, asiago, pomfret, mealworm, emu bone free, and a few others) were left out. See Supplementary Table 3 and Supplementary Table 4 for a longer list of correctly mapped (most popular and rarely used) ingredients and those left unmapped (Supplementary Table 5) with the BERT-based strategy.

**Carbon footprint of food ingredients**

The integration of SuEatable with RecipeDB presents an interesting perspective on the ecological impact of food ingredients and their popularity in recipes (Figure 1d). Logarithmic scales were used to depict ingredient popularity (log 10) and carbon footprint (log 2) due to the wide range of data and uneven scatter. Some ingredients are highly prevalent and are used in thousands of recipes across cuisines (onion, 69096; butter, 54026; garlic clove, 49786), whereas many are rarely used. Carbon footprints (measured in $CO_2$ equivalent per kilogram) values range between 0.11 (bean and its varieties) and 27.72 (beef and its varieties). Due to the low spatial resolution of data, CF values represent global averages.

Ingredients popularly used in culinary preparations (>1% of recipes) and having a low carbon footprint (CF<1) are shown in 'green color.' These 98 ingredients with low environmental consequences and high culinary prevalence include onion, garlic, water, sugar, tomato, lemon, carrot, ginger, and potato, among others. On the other hand, among the 29 ingredients with serious negative environmental consequences due to heavy culinary use and high carbon footprint (Red: CF>4 and >1% of recipes) include butter, milk, cream, cheeses, varieties of beef, chocolates, vanilla, bacon, lamb, pork, fish, and shrimp. The analysis also revealed 1793 ingredients with severe environmental consequences and low cultural adoption (Orange: CF>4 and <0.1% of recipes). These infrequently used culinary ingredients best avoided for positive environmental outcomes include lobsters and variations of beef, pork, bacon, and lamb.

**Carbon footprint of ingredient categories**

Ingredients are classified to represent their source, culinary use, and flavor profiles. Accordingly, ingredients were grouped into 27 categories (vegetable, legume, meat, fish, seafood, dairy, spice, etc.), borrowing the RecipeDB's classification system (See Supplementary Table 6). The ingredients' category-wise CF statistics (median, lower, and upper quartiles) present their typical environmental impact (Figure 2a). Ingredients of vegetable, dairy, meat, miscellaneous, herb, and spice categories are among the most frequently used for culinary preparations. The categories with the highest overall CF are meat (10.347), seafood (10.089), dairy (6.827), dish (5.62), and fish (4.154).

Superimposing ingredient popularity in culinary preparations on top of CF data presents a more transparent picture accounting for cultural influences (Figure 2b). With their heavy use in culinary preparations (76.77% of all recipes), dairy products emerge as environmentally the most damaging. Following closely, meat products come next on the list with undesirable environmental consequences, with 71.41% of all recipes using ingredients such as variations of beef, pork, bacon, eggs, and chicken. Despite high CF, seafood, dish, and fish ingredients are infrequently used in recipes (6.64%, 4.98%, and 3.19%, respectively).

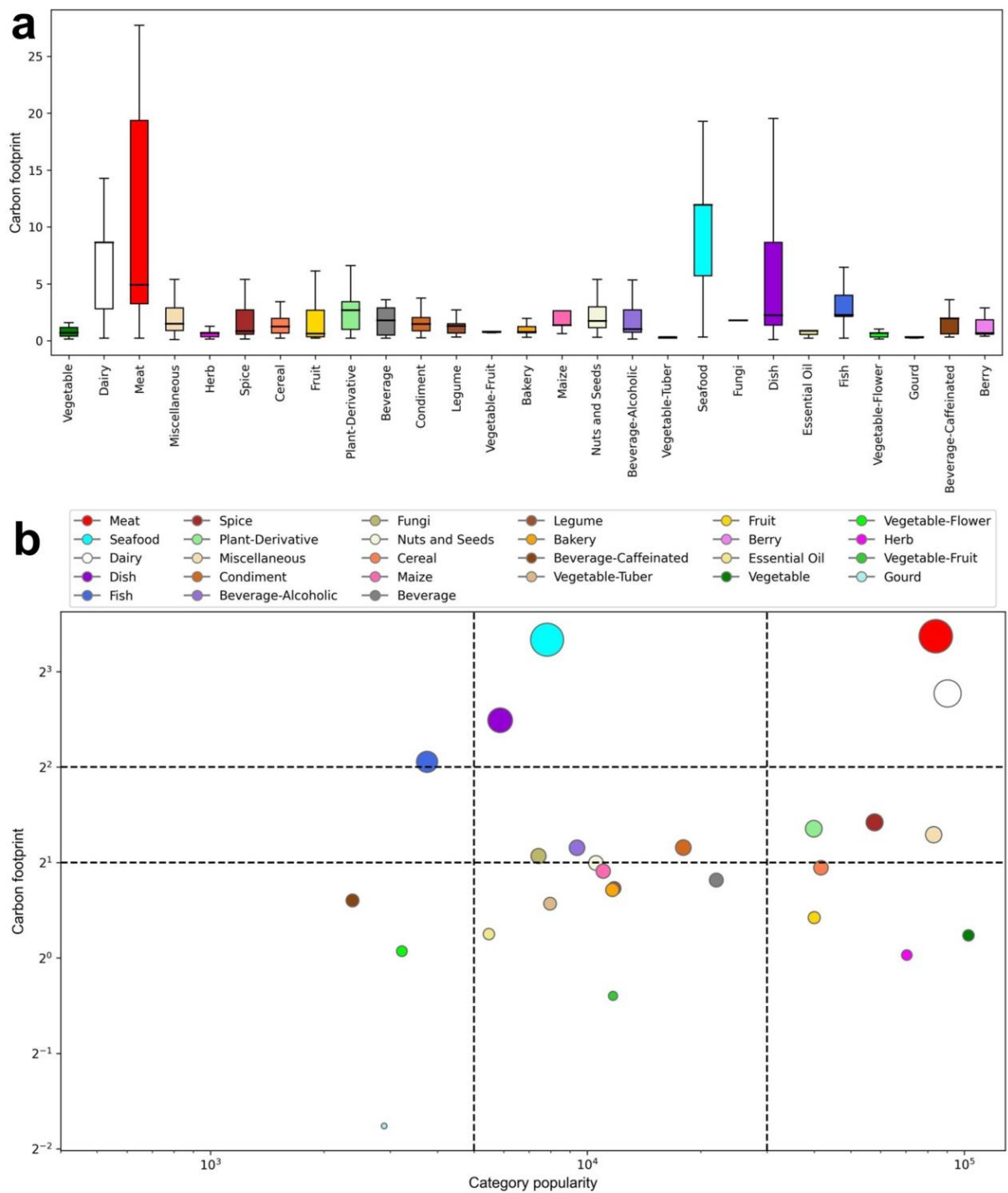

**Figure 2: Carbon footprint analysis for ingredient categories.** (a) CF statistics of ingredient categories (ordered according to decreasing number of ingredients). The box plot shows the median, first and third quartiles, and min-max of data. (b) Environmental impact of various ingredient categories and their popularity in culinary preparations. The bubble size is proportional to the average CF of ingredients in the category.

Vegetable, herb, and fruit categories of ingredients are of high culinary utility (87.21%, 59.82%, and 33.99% of recipes, respectively) with comparatively low emissions (1.178, 1.021, and 1.340, respectively). Incidentally, gourd (cucumber), vegetable flower (broccoli and cauliflower), and caffeinated beverage (cocoa, coffee powder) classes have rare culinary usage (2.45%, 2.73%, and 2.02% of recipes) and the lowest footprint (0.295, 1.05, and 1.518, respectively).

**Estimating carbon footprint of recipes**

Recipes are the cultural capsules that dictate dietary consumption. Beyond the level of ingredients and their categories, the actual environmental impact of food is better assessed by estimating the carbon footprint of recipes, the basic units of dietary consumption, the metaphorical culinary currency. By mapping ingredients in recipes to their global average carbon footprint values, we estimated a ballpark figure of their carbon footprint. 'Recipe Coverage' captures the proportion of all the recipe ingredients mapped with a SuEatable food item. A 100% recipe coverage suggests none of its ingredients were left out in the process of mapping and, hence, is the ideal situation when estimating the recipe's carbon footprint. Good recipe coverage leads to reasonable CF estimates.

Using exact match and BERT-based mapping protocol (as discussed earlier in the section on 'Integration of SuEatable & RecipeDB,' Figure 1c), we achieve good coverage across all the recipes (Figure 3a). Most recipes (99.62%) could be included in the analysis with a lenient recipe coverage expectation of ≥10%. Even with a stringent recipe coverage cutoff of ≥50%, a significant number of recipes were accounted for (88.47%), indicating the effectiveness of the strategy implemented for mapping the ingredients. Given the heterogeneous distribution of the ingredient popularities, it is pertinent that the mapping protocol does not miss the most frequently used ingredients (Figure 3b). The BERT-based embedding maps 71% and 67.5% among the top 100 and 200 most popular ingredients, respectively (Extended Data Figure 2). On average, half of all ingredients in each recipe were matched with their carbon footprint while primarily missing infrequently used ingredients (Figure 3c).

**Comparing carbon footprints of cuisines**

The cuisine-level CF statistics present an interesting picture of the environmental impact of culinary practices from various geo-cultural regions (Figure 4a). The carbon footprint of all the recipes across the 26 cuisines (World) varies between 0.15 and 166.019, with a median CF of 15.64. For example, among the recipes with low carbon footprints are the Mexican dish 'Cilantro-Lime Rice' and the French dish 'Sobronade' with estimated CF of 0.44 and 1.88, respectively. Towards the higher end of the spectrum are the Italian dish 'North End Sunday Gravy' (116.84) and the US dish 'Ring of Fire Chili' (166.02).

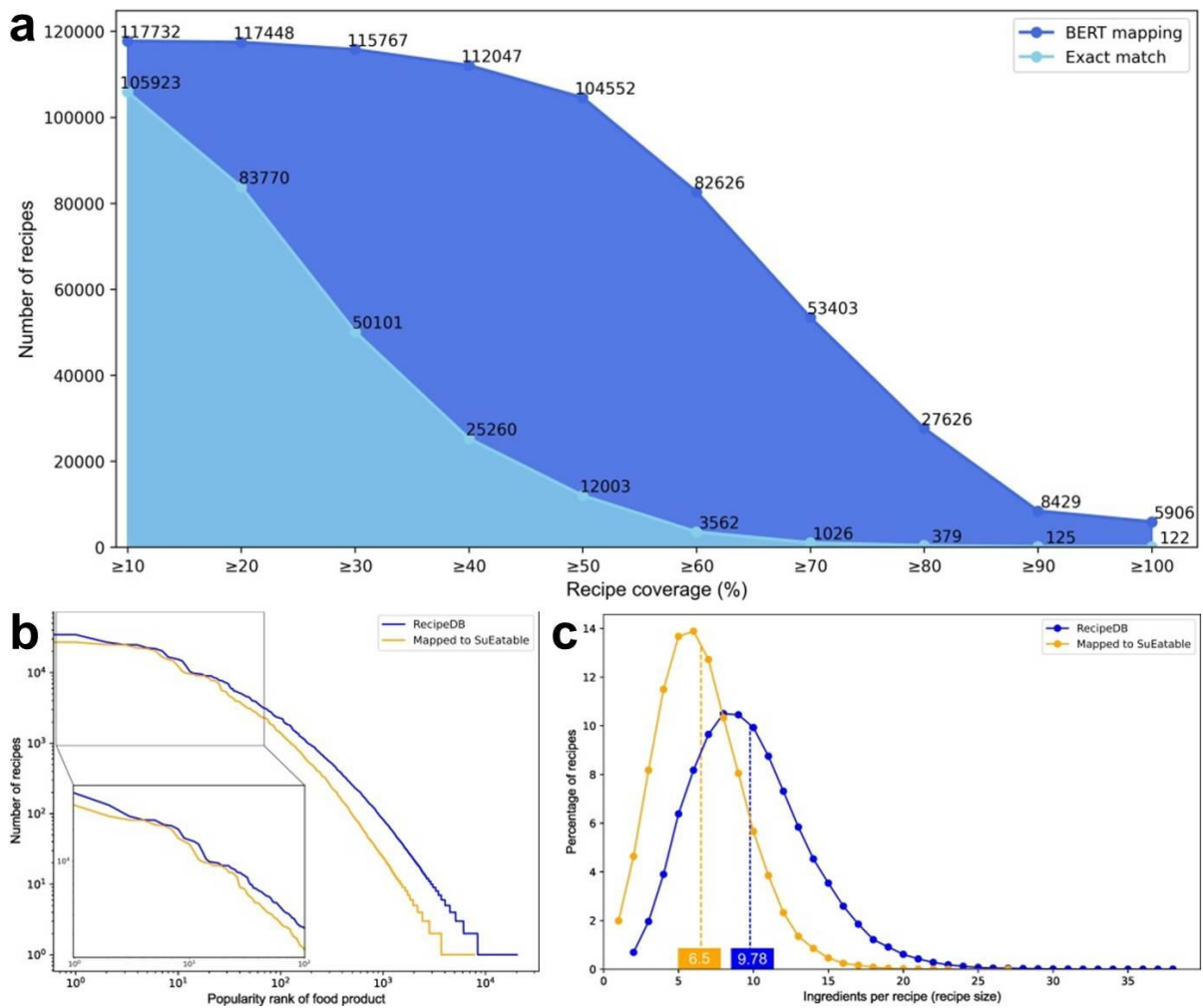

**Figure 3: Effectiveness of the BERT-based strategy for mapping ingredients from recipes to their carbon footprints.** (a) Recipe coverage statistics present the number of recipes with a coverage above a threshold. (b) Comparative statistics show that very few of the most popular ingredients were left out of the mapping protocol. The inset zooms in on Top200 ingredients to highlight the success of the BERT-based mapping. (c) Recipe size distribution, before and after mapping. The average values are highlighted for each of the distributions.

Dishes such as the Chinese 'Rosemary Rice' (17.21) are in the middle of the spectrum. Cultural practices that influence the idiosyncratic ingredient usage patterns render recipes in some cuisines more environmentally harmful than the global average. Among the cuisines with higher median CF than the global median are South American, Belgian, French, Italian, Eastern European, Irish, Scandinavian, Deutschland, UK, US, Greek, and Central American. On the other hand, ingredient combinations in the recipes of the Indian Subcontinent, Caribbean, Korean, Japanese, Thai, Middle Eastern, Rest Africa, Chinese and Mongolian, Southeast Asian,

Spanish and Portuguese, Northern Africa, Mexican, Canadian, and Australian cuisines render them environmentally less harmful with a lower median CF lower than the global value.

Probing deeper into the CF distribution of recipes in each cuisine presents better insights (inset of Figure 4b). Carbon footprint distribution indicates a bounded, Gaussian-like distribution for all 26 cuisines. Most cuisines had high CF recipes above the global average. The cumulative CF distribution helps discern the cuisine-to-cuisine differences (Figure 4b). Recipe size cannot explain these differences, as cuisines have comparable recipe sizes on average (Extended Data Figure 3). Interestingly, the observed differences in the carbon footprints of the cuisines are best explained by the heavy use of high-CF food products (Extended Data Figure 4, Supplementary Figure 2, Supplementary Figure 3, and Extended Data Figure 5).

**Vegetarian and non-vegetarian recipes**

Going beyond investigating the carbon footprints of food products at the level of ingredients, their categories, recipes, and cuisines, we further probed the sustainability of vegetarian and non-vegetarian recipes. The prevailing literature pins the onus of the observed carbon footprint of the food system primarily on animal-sourced products[7]. Heavy consumption of animal-derived products could be one basis of such inference, other than inherent emissions linked to the product. Culturally influenced culinary differences shape the composition of recipes, dictating the use of animal products and other ingredients that have a bearing on the recipe CF. To probe the role of cultural influences in specifying the carbon load of recipes from various cuisines, we segregated the recipes into those using animal-derived food products and those that do not. A recipe was classified as 'non-vegetarian' if it had one or more animal products of either the 'meat,' 'seafood,' or 'fish' category. Among the remaining recipes, those incorporating at least one ingredient from the following categories were labeled as 'vegetarian': vegetable, vegetable-tuber, fruit, plant derivative, legume, nuts, seeds, maize, and vegetable-flower. Miscellaneous recipes exclusively use ingredients that are neither from 'vegetarian' nor 'non-vegetarian' categories.

Overall, the collection of global recipes is dominated by non-vegetarian recipes, with 64.33% of culinary preparations incorporating at least one animal product (Extended Data Figure 6). Some cuisines are significantly enriched with non-vegetarian recipes compared to the world average - South American, Italian, Greek, Central American, French, Irish, Belgian, Eastern European, UK, Scandinavian, and Deutschland. On the other hand, some others have a lesser proportion of recipes that include animal-sourced ingredients - Thai, Japanese, Indian Subcontinent, Chinese and Mongolian, Caribbean, Southeast Asian, Korean, Spanish and Portuguese, Rest Africa, Canadian, US, Australian, Middle Eastern, North Africa, and Mexican.

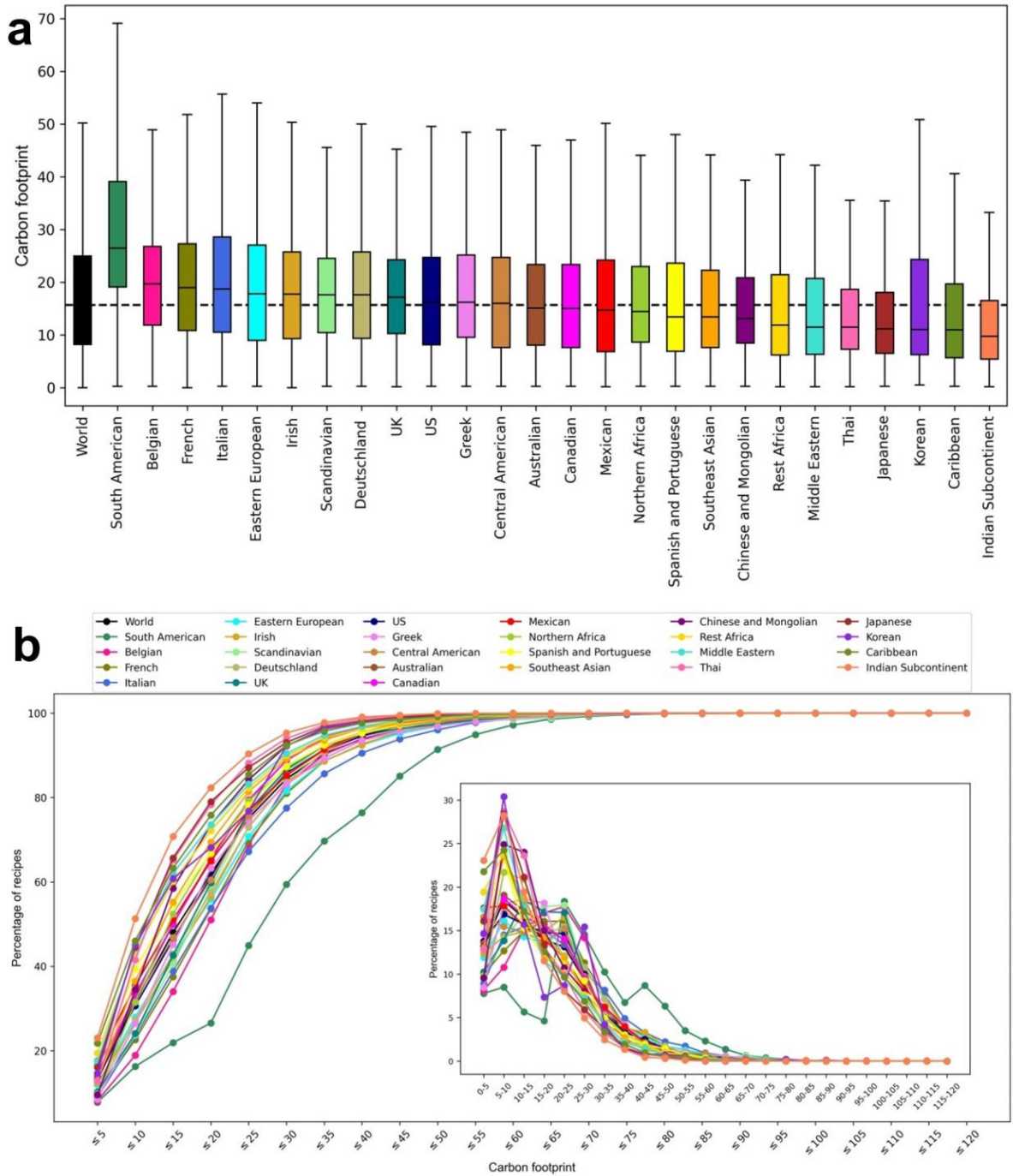

**Figure 4: Comparison of recipe carbon footprints across world cuisines.** (a) Recipe CF statistics across cuisines (ordered according to decreasing value of median CF). The box plot shows the median, first and third quartiles, and min-max of data. (b) The cumulative distribution of recipe CF of cuisines helps better segregation of cuisines. The inset shows recipe CF distributions.

**Carbon footprint of vegetarian and non-vegetarian recipes**

The average carbon footprint of recipes that use meat, egg, fish, or seafood is 22.15 (Standard Deviation: 12.84) as compared to significantly lower (Kolmogorov-Smirnov Test) environmental impact of vegetarian recipes (10.638, Standard Deviation: 7.94) devoid of such animal-derived ingredients (Figure 5a). A consistent pattern across all 26 cuisines suggests that cultural influences render non-vegetarian recipes environmentally unsustainable (Figure 5b). Among the cuisines with a carbon load of non-vegetarian recipes above the global average are South American, Italian, Greek, Central American, and French. Mexican, Northern Africa, and Middle Eastern recipes are at par. Thai recipes have an exceptionally low carbon footprint. Among the other cuisines with recipes employing animal-sourced ingredients that are most environment friendly include Korean, Chinese and Mongolian, South American, Indian Subcontinent, Middle Eastern, Southeast Asian, and Northern Africa.

Idiosyncratic ingredient combinations in vegetarian recipes can make them environmentally untenable despite the exclusive use of plant-derived products. While not using animal products, counter-intuitively, the biased combinations of high CF plant ingredients leave some cuisines (Italian, Belgian, and UK) with a heavy carbon load of vegetarian recipes. Korean, Southeast Asian, Caribbean, Spanish and Portuguese, Thai, South American, Japanese, and Indian Subcontinent cuisines are salient among those with the lowest footprint of vegetarian recipes, as expected.

**Webserver implementation**

We have created a web server (SustainableFoodDB) to enable the exploration of the carbon footprints of recipes using their titles or ingredients used. The search page provides a vegetarian and non-vegetarian label of recipes other than their estimated carbon footprints. Recipe-specific pages provide a comprehensive picture along with the cuisine of the recipe, its constituent ingredients, and the availability status of their CF values. Ingredient-specific pages provide the source(s) of data[4] for their carbon footprints. The server also features a 'Carbon Footprint Calculator', which facilitates computation of estimated carbon footprints of recipes in a user-friendly manner when fed with the list of ingredient and their quantities. The web server implementation was done using a tech stack of ReactJS, CSS, Bootstrap, and MaterialUI for the front end and MongoDB, NodeJS, and ExpressJS for the back end. SustainableFoodDB: https://cosylab.iiitd.edu.in/SustainableFoodDB/.

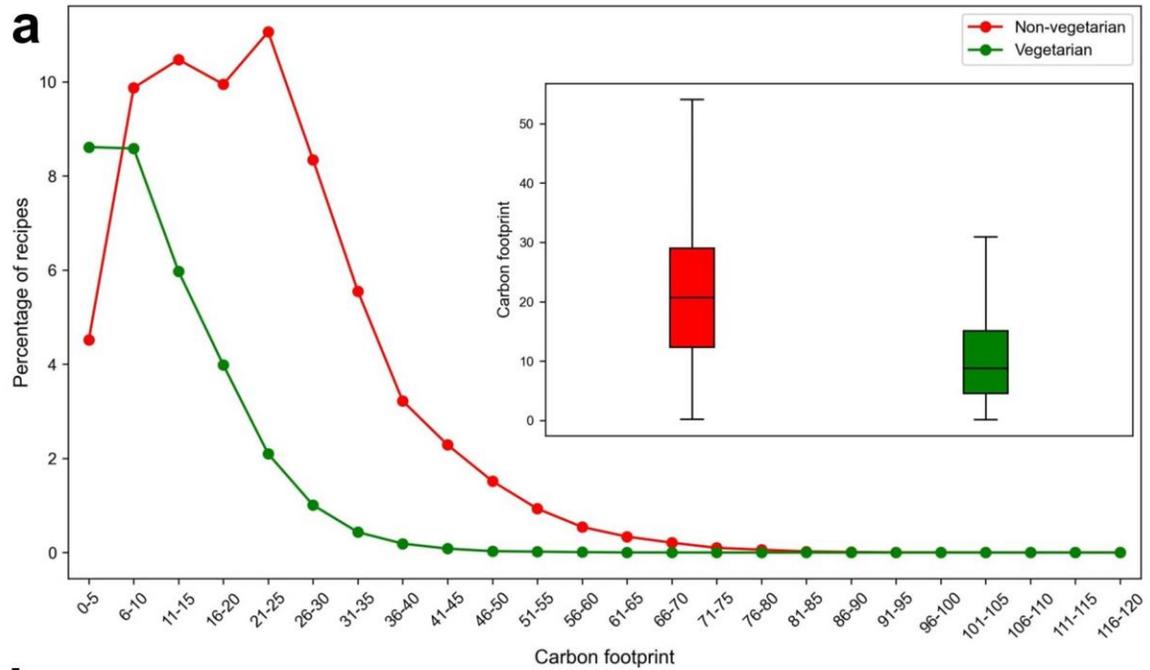
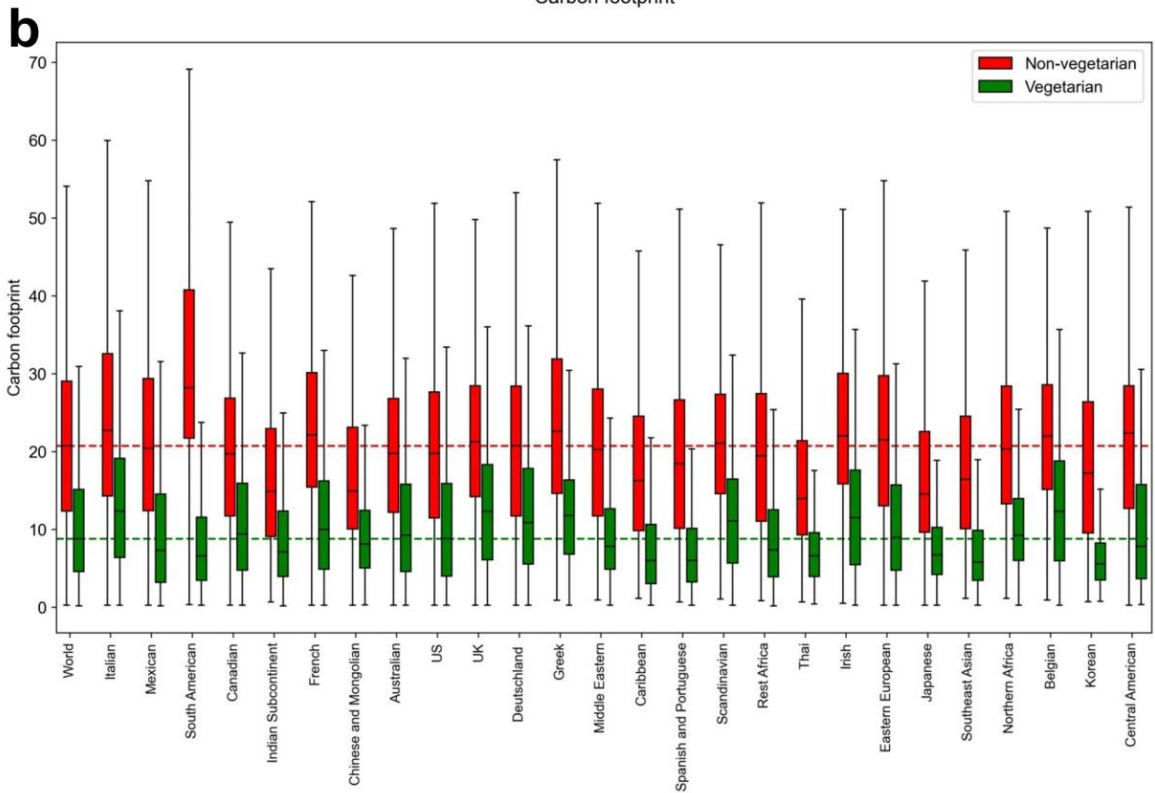

**Figure 5: Comparison of the environmental impact of vegetarian and non-vegetarian recipes.** (a) Carbon footprint distribution of vegetarian and non-vegetarian recipes (World). The comparative statistics of average carbon footprints. (b) Comparison of the median CF of the vegetarian and non-vegetarian recipes across the cuisines. The box plot shows the median, first and third quartiles, and min-max of data.

**Discussion**

Dietary choices are dictated by various intertwined factors, including taste specified by ingredient combinations[28–31], nutrition[32,33], cost, allergies[34], religion[35], and psychology[36]. These choices may have serious deleterious consequences for the environment due the GHG emissions arising from the food system[1]. For a sustainable food ecosystem, we need concerted efforts to collect data on all aspects of the food system. Quantification of the carbon footprint of dietary choices will enable mitigation of the adverse environmental impacts due to global warming and climate change[2,3]. By blending the state-of-the-art carbon footprint data of food products[4] and recipe composition[6], we systematically investigate the carbon load of traditionally consumed recipes and compare cuisines for their environmental impact. We also present an extensive repository[13] of estimated carbon footprints of recipes from across the world. Among the key results, recipe composition has a strong bearing on its carbon load, and recipes using animal-based products tend to have a higher footprint than their plant-based counterparts[7].

While this study captures the estimated carbon footprint of popularly consumed recipes by plugging in CF data from an extensive list of ingredients, going forward, it is desirable to account for the quantity of the ingredients used. Further, beyond the global estimates, the carbon footprint data needs to be at higher geospatial resolution. Various factors contribute to the footprint of food products such as transportation mechanisms[21,22] and consumption patterns[9]. The carbon footprint focuses primarily on greenhouse gas emissions[18], and many factors, such as water usage, land degradation, and resource depletion, may not be fully accounted for.

With increasingly nuanced food labels introduced by the regulatory bodies, carbon footprint values could potentially be used to highlight environmental impact, along with the nutrition and allergy indications. Among the most exciting directions in artificial intelligence is emulating culinary creativity using large language models to generate novel recipes[11]. Data-driven strategies can therefore be leveraged to generate hitherto unseen recipes that are not only tasty but also environmentally sustainable. The adverse impacts of animal-sourced ingredients indicates that efforts towards producing industrial-scale plant-based meat and lab-grown meat are timely and relevant.

Livestock management is one of the dimensions with much scope for implementing strategies to reduce emissions such as carbon sequestration via improved pasture management and better livestock integration in the circular bioeconomy[37]. Reducing the CF of the food system is a much broader challenge and would require systemic change in dietary behaviors, adoption of energy-efficient technologies, transition to renewable energy sources, mechanisms for food waste decarbonization[38], and innovations in agricultural practices, food processing techniques, and transportation technologies. Aligned with the idea of computational gastronomy[24], quantification of carbon footprints of recipes through the lens of cultural context will go a long way in steering the food system towards sustainability.

## Methods

### Manual data curation

SuEatable[4] dataset was cleaned to address typos, bring consistency in country and food item names, achieve data structure consistency, implement stemming, and purge duplicate entries. The ingredient names in RecipeDB[6] were obtained using state-of-the-art named entity recognition algorithms[39,40]. The following changes were made in the food items names: 'chocolate or cream filled cookies' to 'chocolate cream cooky', 'simple cookies' to 'cooky', 'mineral water' to 'water', 'beef bone free meat' to 'beef bone free', 'beef meat with bone' to 'beef with bone', 'chicken bone free meat' to 'chicken bone free', 'chicken meat with bone' to 'chicken with bone.' Similar changes were done for lamb and pork ingredient names.

### Mapping recipe ingredients to carbon footprints

The embeddings for RecipeDB ingredients and SuEatable food products were obtained using Word2Vec[25], RoBERTa[26] (roberta-base-nli-mean-tokens) and BERT[27] (bert-base-nli-mean-tokens) models. The cosine similarity metric was used to obtain the closet match for an ingredient name with its food product with a cutoff of >81%.

### Carbon footprint comparison of cuisines

Pearson's correlation coefficient was used when associating the median CF of cuisines with the presence of high-CF ingredients in recipes and those among the most popular ingredients of the cuisine.

### Comparing the carbon footprint of vegetarian and non-vegetarian recipes

Kolmogorov-Smirnov test (with an alpha value of 0.05) was implemented to find the statistical significance of the difference between the carbon footprints of vegetarian and non-vegetarian recipes. The null hypothesis was, 'carbon footprint distributions of vegetarian and non-vegetarian recipes are indistinguishable.'

**Acknowledgments**

GB thanks Indraprastha Institute of Information Technology Delhi (IIIT-Delhi) for the computational support. GB thanks Technology Innovation Hub (TiH) Anubhuti for the research grant. MG is a research scholar in the Complex Systems Laboratory and is thankful to IIIT-Delhi for the research fellowship. This study was supported by the Infosys Center for Artificial Intelligence, IIIT-Delhi. The authors thank Shreeyash Khalate, Shristi Kotaiah, Pranshu Patel, and Palak Kasoundhan for implementing the first version of the webserver.

**Author contributions**
GB conceptualized the idea, designed the methodology, and supervised the project. MG, VN, SD, and VK preprocessed and curated the SuEatable dataset. MG conducted the analysis and visualized the results. SS implemented the web server. GB and MG administered the project and wrote the manuscript.

**Competing interest**

The authors declare no competing interests.

**Additional information**

See Supplementary Information.

**Extended Data Figures and Tables**

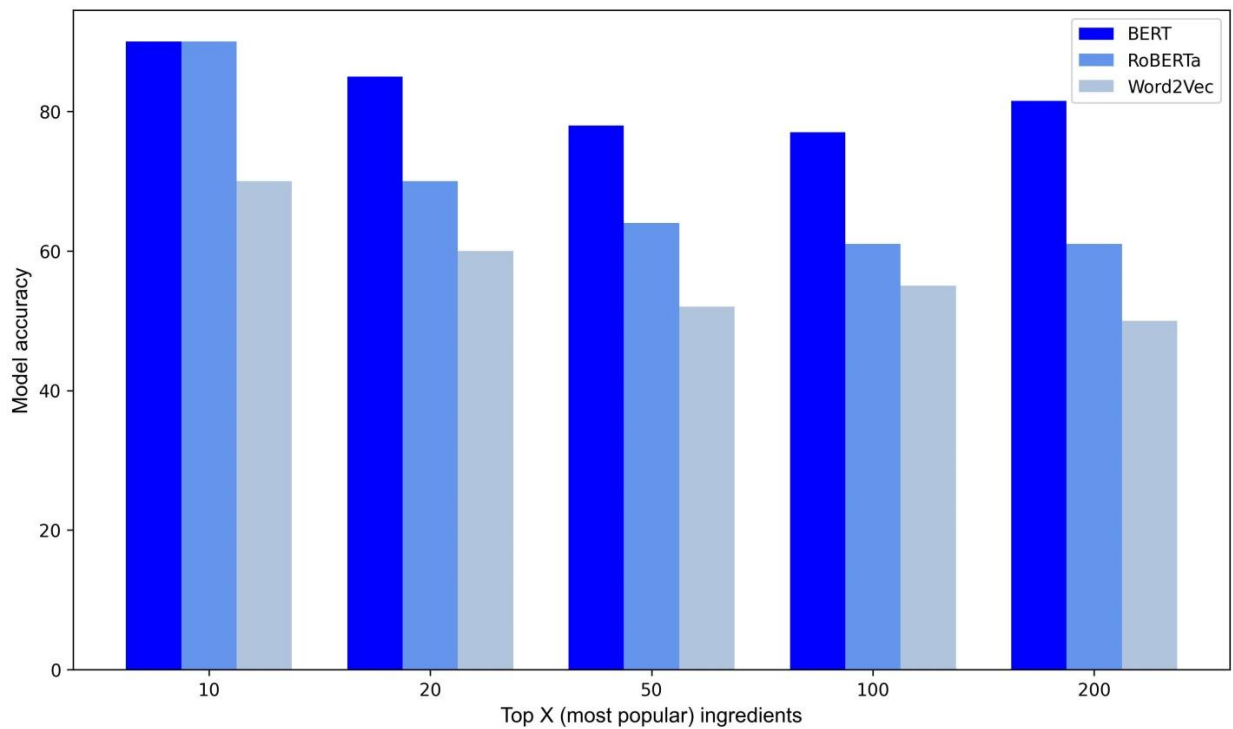

**Extended Data Figure 1: Comparison of performance of models in accurately mapping RecipeDB ingredients to SuEatable food product.** We used word embeddings generated with BERT, RoBERTa, and Word2Vec for mapping ingredients. For each model, the performance was evaluated for mapping 200 most popular RecipeDB ingredients to their SuEatable counterparts by manual assessment. The figure presents model accuracy for the Top 10, 20, 50, 100, and 200 most popular ingredients for all three models. While, expectedly the performance drops with increasing number of ingredients considered, BERT emerged as the best model. The ingredient pairs were pruned by using >81% as the cosine similarity cutoff across all three models.

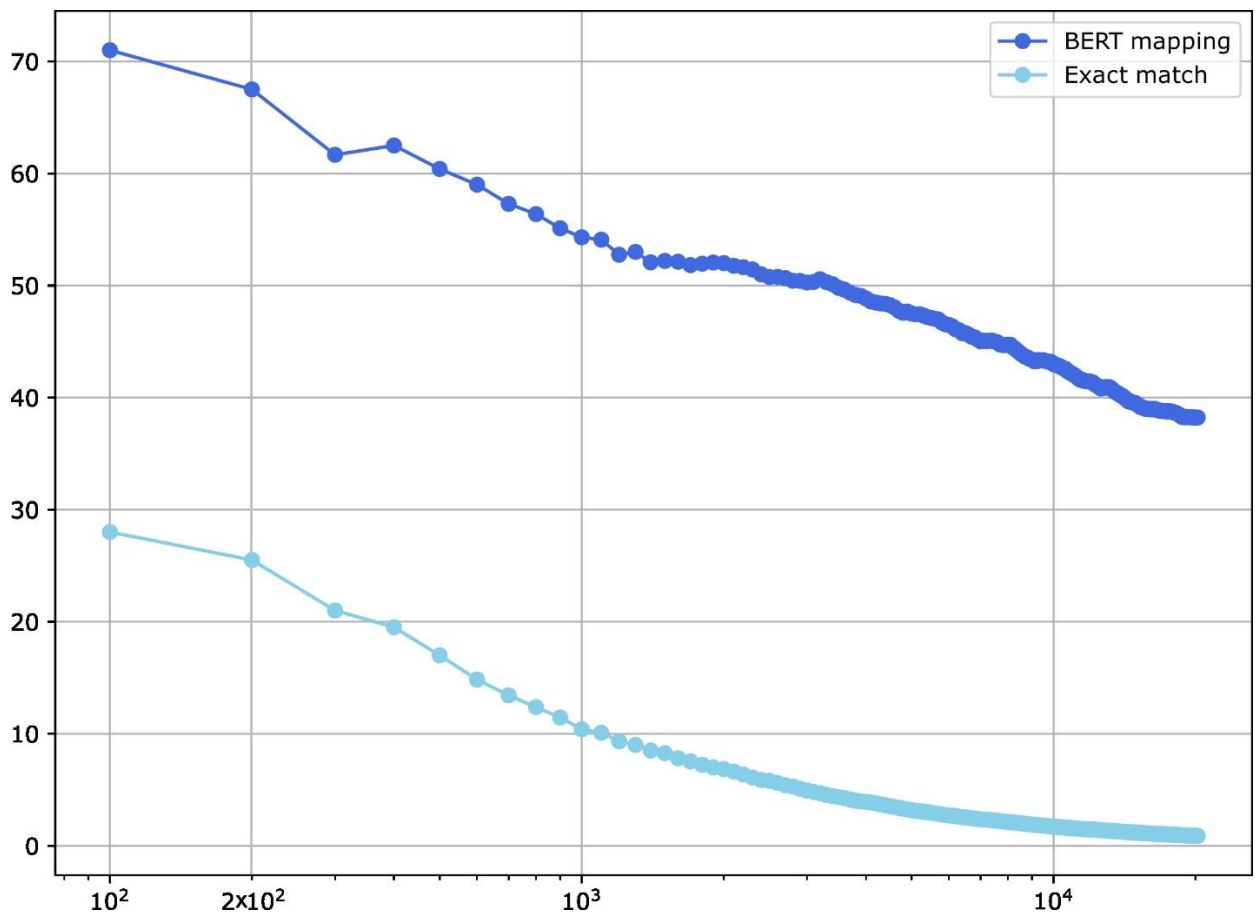

**Extended Data Figure 2: Percentage of RecipeDB ingredients successfully mapped to SuEatable for different cutoffs of popularity (Top X most popular).** These statistics evaluate mapping protocols in matching the most popular ingredients with their carbon footprint. Further to its superior performance (Extended Data Figure 1), the ability of the BERT model in mapping the most popularly used ingredients in recipes makes it the best candidate and hence was used as the basis for further calculations. Please see Figure 1 in the 'Supplementary Information' for a comparison of 'Exact Match' strategy with all the models implemented (BERT, RoBERTa, and Word2Vec).

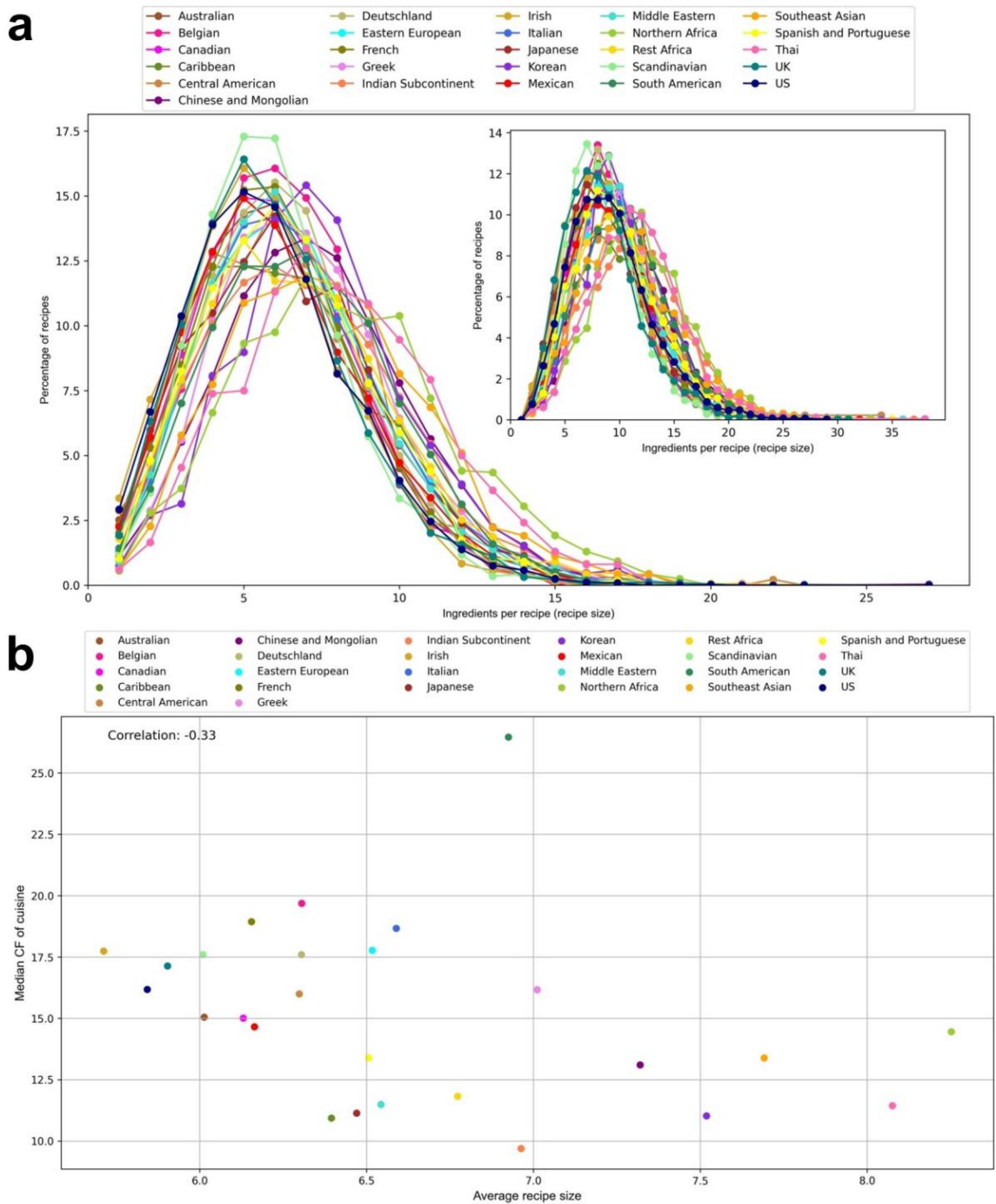

**Extended Data Figure 3: Recipe size statistics and their association with median carbon footprint of cuisines.** (a) This statistics present recipe size distribution of cuisines before mapping ingredients (inset) and after BERT-based mapping. The latter shows reduced average recipe sizes, as SuEatable does not have carbon footprint value for every RecipeDB ingredient. (b) Importantly, the average recipe sizes are comparable across the world cuisines and do not show correlation with their median carbon footprint values.

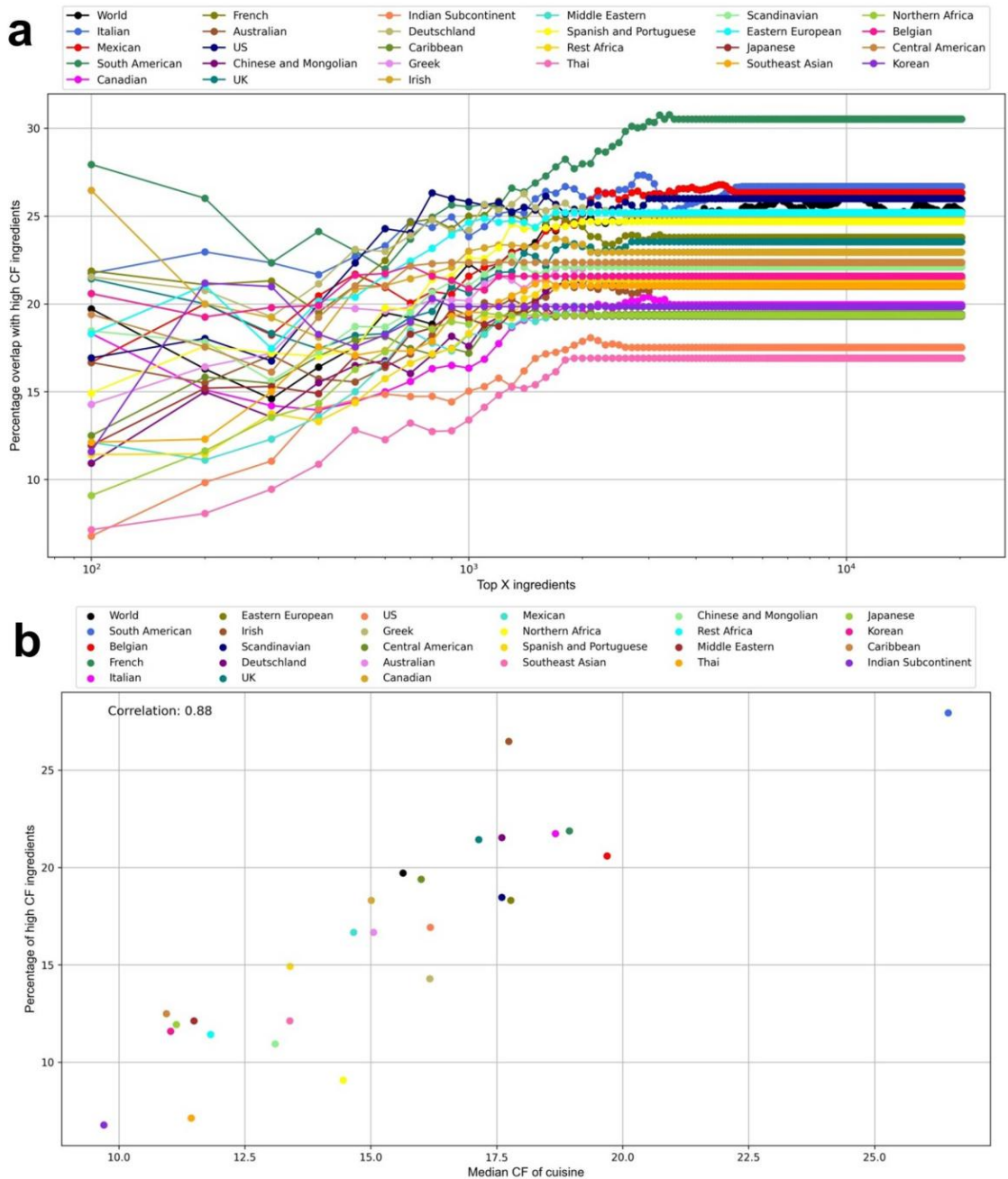

**Extended Data Figure 4: Overlap of Top X RecipeDB ingredients with High CF (CF>4) ingredients.** (a) The number of shared ingredients between the top most popular food products and having severe environmental impact (CF>4). Clearly, the cuisines get segregated based on how many of the most frequently used ingredients in their recipes are environmentally unsustainable. (b) The median CF of cuisines shows a high correlation (0.88) for the Top 100 most frequently used ingredients. Please see Figure 2 and Figure 3 in the 'Supplementary Information' for the correlation of median CFs with the Top 1000 and Top 10,000 ingredients, respectively.

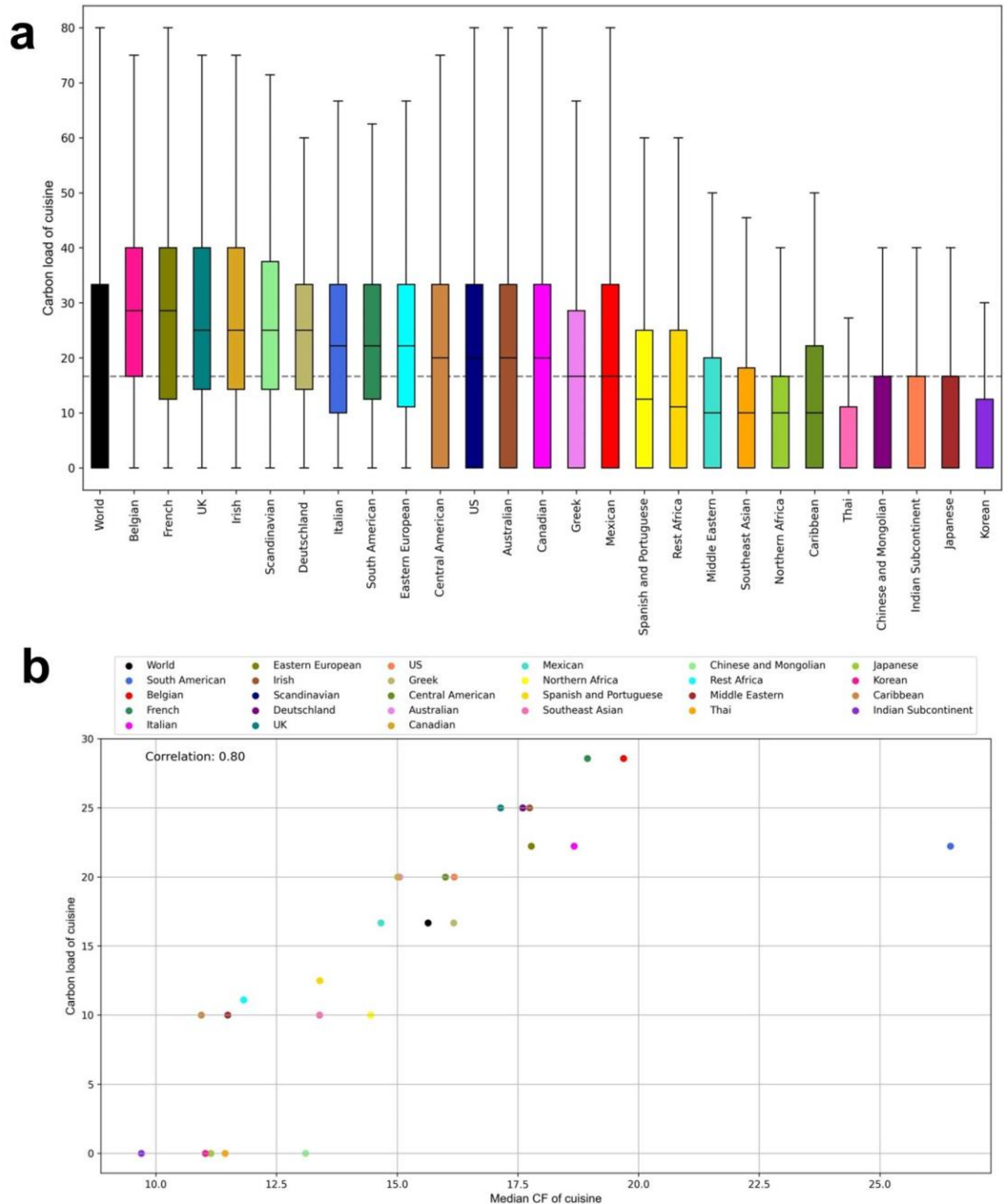

**Extended Data Figure 5: Carbon load of cuisines.** The percentage of high CF ingredients in a recipe is defined as its 'carbon load.' Accordingly, averaging across all the recipes in the cuisine will yield the carbon load of the cuisine. (a) Carbon load of cuisines. The dotted grey line represents the global median. The global average of recipes with high CF ingredients is 19.95% (Median: 16.66%, Standard deviation: 19.18%). The percentage of recipes in Belgian, French, UK, Irish, Scandinavian, Deutschland, Italian, South American, Eastern European, Central American, US, Australian, and Canadian regions are higher than the global median value. (b) The median CF of cuisines shows a high correlation (0.80) with the percentage of high CF ingredients in recipes.

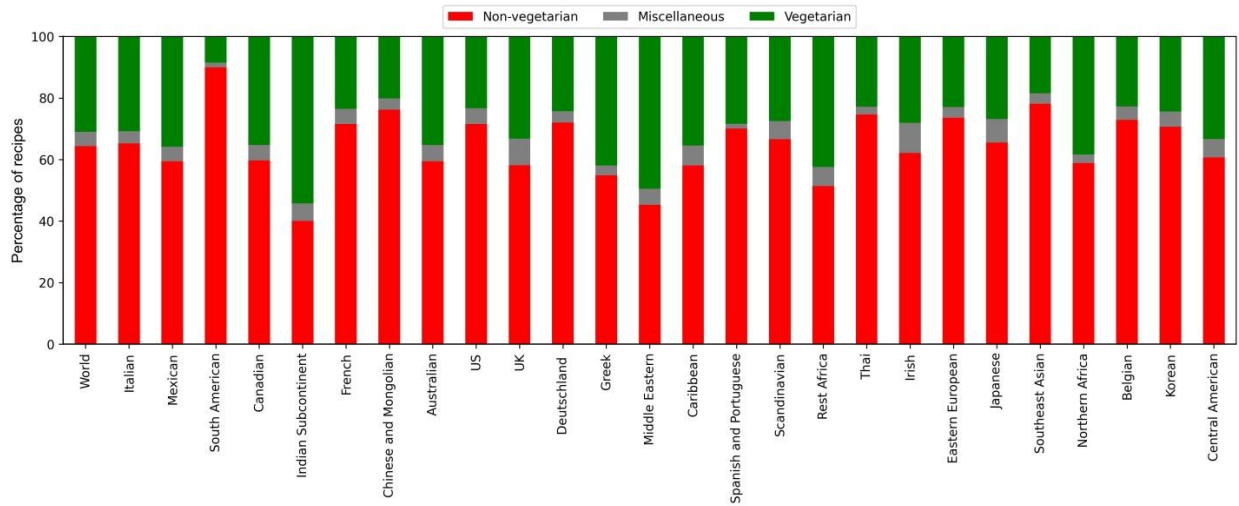

**Extended Data Figure 6: Extent of non-vegetarian recipes.** Statistics of vegetarian, non-vegetarian, and miscellaneous recipes across cuisines.

# SUPPLEMENTARY INFORMATION

**Supplementary Table 1: List of Regions and Countries in the SuEatable dataset.**

| Region | Country |
|---|---|
| Africa | Africa (Near East And North) |
| Africa | Africa (Sub-Saharan) |
| Africa | Ghana |
| Africa | Madagascar |
| Africa | Mauritius |
| Africa | Morocco |
| Africa | North Africa |
| Africa | Senegal |
| Africa | Somalia |
| Africa | South Africa |
| Africa | Tanzania |
| Africa | Uganda |
| Asia | Asia (East And South-East) |
| Asia | Asia (South) |
| Asia | Bangladesh |
| Asia | China |
| Asia | E Asia |
| Asia | India |
| Asia | Indian Ocean |
| Asia | Indonesia |
| Asia | Iran |
| Asia | Japan |
| Asia | Kazakhstan |
| Asia | Korea |
| Asia | Malaysia |
| Asia | Maldives |
| Asia | Pakistan |
| Asia | Philippines |
| Asia | Thailand |
| Asia | Vietnam |
| Atlantic Ocean | Atlantic Ocean |
| C America | Costa Rica |
| C America | El Salvador |
| C America | Guatemala |
| C America | Mexico |
| C America | Nicaragua |
| C Europe | C Europe |
| CS America | America (Central And South) |
| E Europe | Bulgaria |
| E Europe | Czech Republic |
| E Europe | E Europe |
| E Europe | Estonia |
| E Europe | Hungary |
| E Europe | Latvia |
| E Europe | Lithuania |
| E Europe | Romania |
| E Europe | Russia |
| E Europe | Serbia |
| E Europe | Slovakia |
| E Europe | Slovenia |
| Europe | EU |
| Europe | Europe |
| Mediterranean Area | Croatia |
| Mediterranean Area | Cyprus |
| Mediterranean Area | France (Corsica) |
| Mediterranean Area | Greece |
| Mediterranean Area | Israel |
| Mediterranean Area | Italy |
| Mediterranean Area | Malta |
| Mediterranean Area | Mediterranean Area |
| Mediterranean Area | Portugal |
| Mediterranean Area | Spain |
| Mediterranean Area | Spain (Valencia) |
| Mediterranean Area | Tunisia |
| Mediterranean Area | Turkey |
| N America | Alaska |
| N America | Canada |
| N America | Canada (East) |
| N America | Canada (West) |
| N America | N America |
| N America | USA |
| N America | USA Mid-West |
| N Europe | Austria |
| N Europe | Belarus |
| N Europe | Belgium |
| N Europe | Denmark |
| N Europe | Finland |
| N Europe | France |
| N Europe | France (North) |
| N Europe | France (South) |
| N Europe | Germany |
| N Europe | Iceland |
| N Europe | Ireland |
| N Europe | Luxembourg |
| N Europe | Moldavia |
| N Europe | N Europe |
| N Europe | Netherlands |
| N Europe | Norway |
| N Europe | Poland |
| N Europe | Sweden |
| N Europe | Switzerland |
| N Europe | UK |
| N Europe | UK (Scotland) |
| N Europe | Ukraine |
| Non EU Countries | Non EU Countries |
| Oceania | Australia |
| Oceania | New Zealand |
| Oceania | Oceania |
| Pacific Ocean | Pacific Ocean |
| S Africa | Reunion Island |
| S America | Argentina |
| S America | Brazil |
| S America | Chile |
| S America | Colombia |
| S America | Ecuador |
| S America | Peru |
| S America | S America |
| S America | S America (Latin America And Caribbean) |
| S America | Uruguay |
| S America | Venezuela |
| W Europe | W Europe |
| World | World Estimate |

**Supplementary Table 2: Performance comparison for models implemented to map the embeddings of ingredient names (RecipeDB) to those of food products (SuEatable).**

| Top X Ingredients | Model | Accuracy | Precision | Recall | F1 Score |
|---|---|---|---|---|---|
| 10 | **BERT** | **90** | **100** | **88** | **93.61** |
| | RoBERTa | 90 | 100 | 88 | 93.61 |
| | Word2Vec | 70 | 100 | 66 | 79.51 |
| 20 | **BERT** | **85** | **88.2** | **93.7** | **90.86** |
| | RoBERTa | 70 | 76.47 | 86.66 | 81.24 |
| | Word2Vec | 60 | 100 | 52.94 | 70.53 |
| 50 | **BERT** | **78** | **88.23** | **81.08** | **84.5** |
| | RoBERTa | 64 | 82.97 | 77.14 | 74.99 |
| | Word2Vec | 52 | 100 | 38.46 | 55.55 |
| 100 | **BERT** | **77** | **84.5** | **83.33** | **83.91** |
| | RoBERTa | 61 | 67.56 | 76.92 | 71.93 |
| | Word2Vec | 55 | 100 | 40 | 57.14 |
| 200 | **BERT** | **85.5** | **93.43** | **83.91** | **86.32** |
| | RoBERTa | 61 | 67.85 | 74.21 | 70.88 |
| | Word2Vec | 50 | 100 | 37.41 | 54.45 |

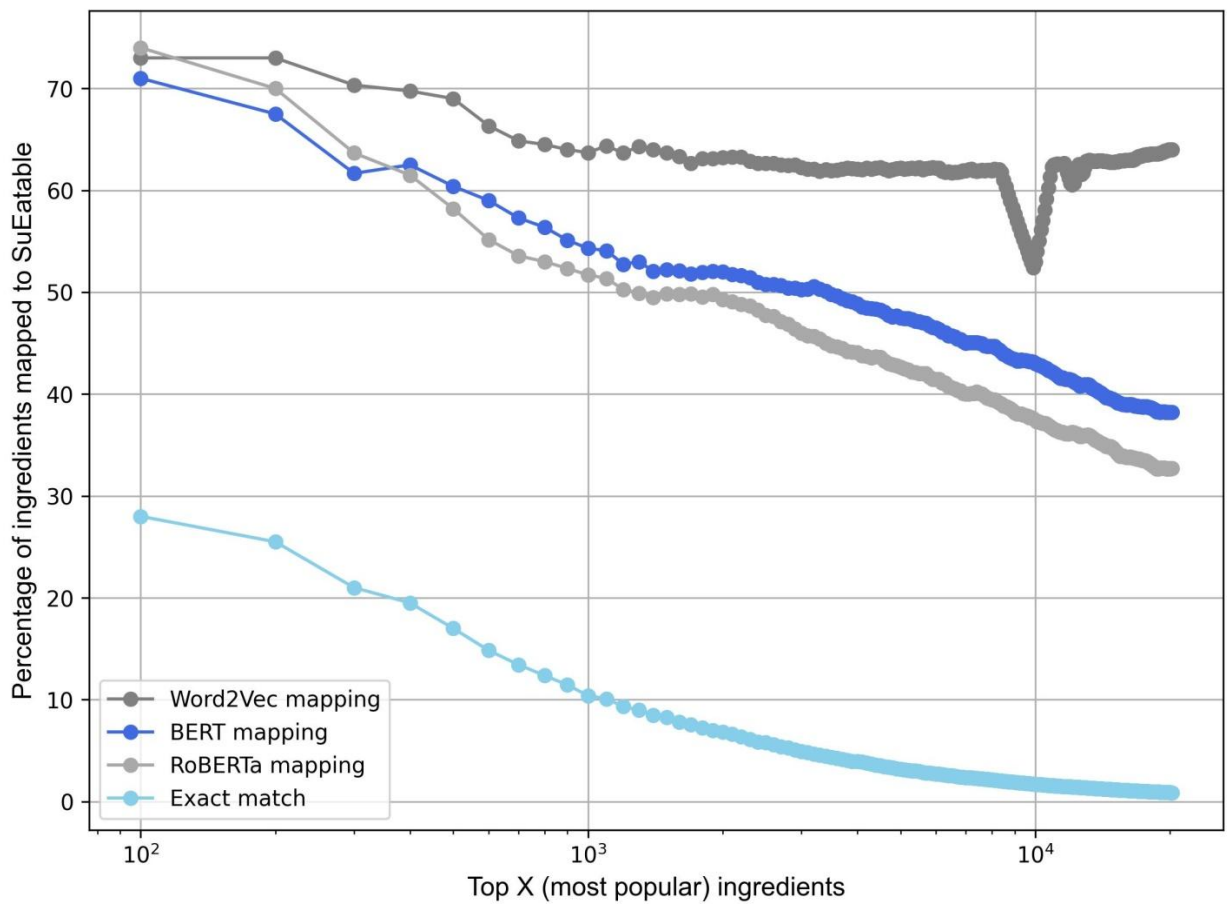

**Figure 1: Comparison of the model performance for successfully mapping Top X RecipeDB ingredients to SuEatable.** Despite an apparently superior mapping of ingredients of Word2Vec model, by comparing with the manually curated ground truth, we show that BERT is a better model (See Supplementary Table 2).

**Supplementary Table 3: 50 most popular RecipeDB ingredients and their SuEatable mappings (Accuracy of BERT model for Top 50 ingredients = 78%).**

| RecipeDB Ingredient | SuEatable Food Product | Similarity | CF | Frequency (Number of Recipes) |
|---|---|---|---|---|
| Onion | Onion | 1 | 0.24 | 69096 |
| Butter | Butter | 1 | 9.9 | 54026 |
| Garlic Clove | Garlic | 0.93 | 0.67 | 49786 |
| Water | Water | 1 | 0.49 | 49546 |
| Olive Oil | Olive Oil | 1 | 3.84 | 44782 |
| Egg | Egg | 1 | 3.23 | 43722 |
| Sugar | Beet Sugar | 0.85 | 0.89 | 40542 |
| Tomato | Tomato | 1 | 0.48 | 33250 |
| Garlic | Garlic | 1 | 0.67 | 30872 |
| Milk | Cream | 0.81 | 5.34 | 29170 |
| Pepper | Pepper | 1 | 0.58 | 24608 |
| Salt Pepper | Pepper | 0.82 | 0.58 | 20774 |
| Flour | Millet Flour | 0.88 | 1.37 | 19674 |
| Cinnamon | Vanilla | 0.81 | 4.3 | 19248 |
| Lemon Juice | Lemon | 0.87 | 0.22 | 19018 |
| Carrot | Carrot | 1 | 0.23 | 18174 |
| Purpose Flour | Millet Flour | 0.84 | 1.37 | 18046 |
| Vegetable Oil | Vegetable | 0.89 | 0.69 | 18036 |
| Cumin | Quorn | 0.91 | 2.5 | 17984 |
| Cream | Cream | 1 | 5.34 | 17160 |
| Ginger | Ginger | 1 | 0.88 | 16816 |
| Parmesan Cheese | Cheese Semihard | 0.89 | 8.65 | 15578 |
| Soy Sauce | Soy Cream | 0.92 | 1.62 | 15454 |
| Beef | Beef With Bone | 0.87 | 19.54 | 15348 |
| Potato | Potato | 1 | 0.25 | 14528 |
| Green Onion | Green Bean | 0.86 | 0.73 | 14148 |
| Chicken Broth | Chicken With Bone | 0.87 | 3.25 | 11826 |
| Lemon | Lemon | 1 | 0.22 | 10986 |
| Lime Juice | Lime | 0.81 | 0.22 | 10956 |
| Chicken Breast | Chicken With Bone | 0.90 | 3.25 | 10792 |
| Mushroom | Mushroom | 1 | 1.78 | 10284 |
| Garlic Powder | Garlic | 0.89 | 0.67 | 9532 |
| Celery | Celery | 1 | 0.32 | 9216 |
| Cheddar Cheese | Cheese Semihard | 0.89 | 8.65 | 8996 |
| Cornstarch | Corn Can | 0.91 | 1.36 | 8776 |
| Nutmeg | Cashew Nut | 0.84 | 1.56 | 8486 |
| White Wine | Wine White | 0.93 | 0.74 | 7944 |
| Vanilla Extract | Vanilla | 0.91 | 4.3 | 7912 |

| Honey | Honey | 1 | 1.74 | 7906 |
| Red Bell Pepper | Red Chilli | 0.84 | 0.8 | 7820 |
| Tomato Paste | Tomato Peel | 0.96 | 1.28 | 7636 |
| Coriander | Radish | 0.82 | 0.15 | 7412 |
| Chicken | Chicken With Bone | 0.82 | 3.25 | 7178 |
| Tomato Sauce | Tomato Peel | 0.92 | 1.28 | 6962 |
| Mozzarella Cheese | Cheese Semihard | 0.85 | 8.65 | 6936 |
| Almond | Almond | 1 | 1.9 | 6616 |
| Bacon | Bacon | 1 | 4.03 | 6516 |
| Green Bell Pepper | Green Bean | 0.85 | 0.73 | 6344 |
| Red Pepper | Red Chilli | 0.87 | 0.8 | 6330 |
| Turmeric | Radish | 0.82 | 0.15 | 6216 |
| Cream Cheese | Hazelnut Cream | 0.84 | 2.71 | 6188 |

**Supplementary Table 4: 50 least frequently used RecipeDB ingredients and their SuEatable mappings.**

| RecipeDB Ingredient | SuEatable Food Product | Similarity | CF | Frequency (Number of Recipes) |
|---|---|---|---|---|
| Rainbow Chocolate Chip | Hazelnut Chocolate | 0.91 | 3.43 | 2 |
| Chocolate Candy Melt | Hazelnut Chocolate | 0.90 | 3.43 | 2 |
| Spinach Ravioli | Spinach | 0.83 | 0.48 | 2 |
| Cherry Chocolate Bar | Almond Chocolate | 0.89 | 4.8 | 2 |
| Serrano Pepper Heat | Pepper | 0.84 | 0.58 | 2 |
| Brioche Breadcrumb | Bread Whole | 0.81 | 0.78 | 2 |
| Bread Improver Knead | Bread Multicereal | 0.81 | 0.7 | 2 |
| Abiu | Hake | 0.83 | 10.12 | 2 |
| Chicken Mince | Chicken With Bone | 0.90 | 3.25 | 2 |
| Raspberry Cordial | Raspberry | 0.94 | 0.63 | 2 |
| Fruit Sultana Apricot | Apricot | 0.89 | 0.36 | 2 |
| Cheese Feta | Cheese Semihard | 0.90 | 8.65 | 2 |
| Clix Biscuit | Cucumber | 0.82 | 0.32 | 2 |
| Mango Peach Tea | Peach | 0.85 | 0.43 | 2 |
| Pepper Beef Sausage | Pork Sausage | 0.86 | 17.94 | 2 |
| Cream Preferably | Cream | 0.90 | 5.34 | 2 |
| Chicken Carcass Meat Remaining | Chicken With Bone | 0.81 | 3.25 | 2 |
| Plain Biscuit Cracker Crumb | Plain Cracker | 0.85 | 1.24 | 2 |
| No Oil French Dressing | Pesto Without Garlic | 0.81 | 2.72 | 2 |
| Violet Crumble Chocolate Candy | Almond Chocolate | 0.85 | 4.8 | 2 |
| Bush Tomato | Tomato Peel | 0.86 | 1.28 | 2 |
| Cadbury Chocolate Candy Bar | Hazelnut Chocolate | 0.89 | 3.43 | 2 |
| Hazelnut Milk Chocolate | Hazelnut Chocolate | 0.94 | 3.43 | 2 |
| Lsa | Tangerin | 0.83 | 0.38 | 2 |
| Coffee Creamer Blueberry | Blueberry Juice | 0.86 | 3 | 2 |
| Pumpkin Pepitas | Pumpkin | 0.92 | 0.38 | 2 |
| Bran Cereal Protein Powder | Cornflakes Cereal | 0.81 | 2.64 | 2 |
| Nut Corn Flake | Cornflakes Cereal | 0.88 | 2.64 | 2 |
| Grumichama | Hake | 0.84 | 10.12 | 2 |
| Pear Juice Fruit | Pear Juice | 0.96 | 0.49 | 2 |
| Gluten Flour Mix | Sorghum Flour | 0.88 | 1.33 | 2 |
| Milk Chocolate Dark | Milk Chocolate | 0.86 | 3.6 | 2 |
| Double Espresso | Espresso | 0.87 | 0.55 | 2 |
| Cinnamon Apple Tea | Cranberry Juice | 0.83 | 2.88 | 2 |
| Cheese Tart | Cheese Semihard | 0.90 | 8.65 | 2 |
| Arnott Milk Coffee Biscuit | Coffee Powder | 0.82 | 0.33 | 2 |
| Ciabatta Choice Bread | Bread Whole | 0.81 | 0.78 | 2 |
| Oval Pita Pocket Bread | Bread Multicereal | 0.81 | 0.7 | 2 |
| Cocoa More | Cocoa Cake | 0.82 | 1.97 | 2 |
| Almond One | Almond | 0.95 | 1.9 | 2 |
| Pistachio Paste | Pistachio | 0.93 | 1.6 | 2 |
| Ice Cream Topping Peppermint Crisp | Cranberry Juice | 0.82 | 2.88 | 2 |
| Chicken Schnitzel | Chicken With Bone | 0.85 | 3.25 | 2 |
| Berry Fruit | Pear Juice | 0.81 | 0.49 | 2 |
| Frisee Lettuce | Lettuce | 0.94 | 0.4 | 2 |
| Dutch Cocoa Hershey | Hazelnut Chocolate | 0.82 | 3.43 | 2 |
| Milk Chocolate Hershey | Milk Chocolate | 0.94 | 3.6 | 2 |
| Chocolate Flavor Crisp Rice Cereal | Chocolate Cream Cooky | 0.81 | 1.69 | 2 |
| Chocolate Ripple Ice Cream | Chocolate Cream Cooky | 0.92 | 1.69 | 2 |
| Caramel Topping Syrup | Cranberry Juice | 0.85 | 2.88 | 2 |

**Supplementary Table 5: 60 most popular RecipeDB ingredients that were unmapped to SuEatable**

| | | | |
|---|---|---|---|
| Salt | Cayenne Pepper | Dijon Mustard | Flat Leaf Parsley |
| Black Pepper | Red Pepper Flake | Black Olive | Celery Rib |
| Parsley | Salt Black Pepper | Mustard | Garam Masala |
| Cilantro | Clove | Mint | Pine Nut |
| Oregano | Worcestershire Sauce | Basil Leaf | Mint Leaf |
| Brown Sugar | Kosher Salt | Cumin Seed | Steak |
| Oil | Heavy Cream | Vinegar | Coriander |
| Basil | Baking Soda | Allspice | Sage |
| Extra Virgin Olive Oil | Salsa | White Pepper | Cayenne |
| Baking Powder | Canola Oil | Balsamic Vinegar | Turkey |
| Paprika | Sea Salt | White Vinegar | Ice |
| Bay Leaf | Jalapeno Pepper | White Onion | Rum |
| Chili Powder | Black Bean | Italian Seasoning | Seasoning |
| Thyme | Curry Powder | Jalapeno | Tortilla Chip |
| White Sugar | Rosemary | Ham | Chilli Sauce |

**Supplementary Table 6: List of ingredient categories and associated ingredients.**

| Ingredient Category | Frequency of Ingredient | Illustrative List of Ingredients |
|---|---|---|
| Meat | 1212 | Egg, Beef, Chicken, Bacon, Lamb, Pork, Turkey, etc. |
| Fruit | 724 | Lemon, Raisin, Orange, Coconut, Cherry, Tomato, Apple, Banana, Cranberry, Raspberry, Mango, etc. |
| Dairy | 653 | Butter, Milk, Cream, Cheese, Yogurt, Curd, etc. |
| Cereal | 498 | Flour, Rice, Vermicelli, Green bean, Sesame seed, Pasta, Wheat, Maize, etc. |
| Plant-Derivative | 491 | Olive oil, Honey, Red wine, Vinegar, Chocolate, etc. |
| Dish | 465 | Vegetable broth, Sausage, Tortilla, Tomato soup, Ice-cream, Cake, etc. |
| Vegetable | 464 | Onion, Tomato, Carrot, Spinach, Chilly, Lettuce, Eggplant, Cabbage, Plum, etc. |
| Bakery | 450 | Bread, Biscuit, Tortilla, etc. |
| Miscellaneous | 439 | Water, Sugar, Cinnamon, Vegetable oil, Vanilla extract, Margarine, Ketchup, etc. |
| Beverage | 239 | Lemon juice, Pineapple juice, Green tea, Coffee, Shakes etc. |
| Fish | 223 | Salmon, Tuna, Anchovy, Cod, Haddock, etc. |
| Spice | 198 | Pepper, Salt pepper, Cumin, Ginger, Nutmeg, Turmeric, etc. |
| Legume | 190 | Vanilla, Pea, Chickpea, Bean, lentil |
| Nuts and Seeds | 161 | Almond, Walnut, Peanut, Cashew, Pistachio, Hazelnut |
| Beverage-Alcoholic | 158 | White wine, Red wine, Mirin, Apple cider, Rum, Beer, etc. |
| Condiment | 157 | Soy sauce, Tomato sauce, Mayonnaise, etc. |
| Maize | 146 | Corn and its variations |
| Seafood | 108 | Shallot, Shrimp, Mussel, Prawn, Clam, Scallop, Squid |
| Herb | 105 | Garlic, Coriander, Hummus, Fennel, Thyme, Celery, etc. |
| Fungi | 57 | Mushroom and its variations |
| Vegetable-Flower | 52 | Broccoli, Artichoke, Cauliflower |
| Essential Oil | 49 | Sesame oil, Walnut oil |
| Vegetable-Fruit | 39 | Bell pepper and its variations |
| Beverage-Caffeinated | 39 | Cocoa powder, and Coffee variations |
| Vegetable-Tuber | 31 | Potato and its variations |
| Gourd | 18 | Cucumber and its variations |
| Berry | 7 | Wheat berry, Rye berry, Berry syrup, Berry fruit |

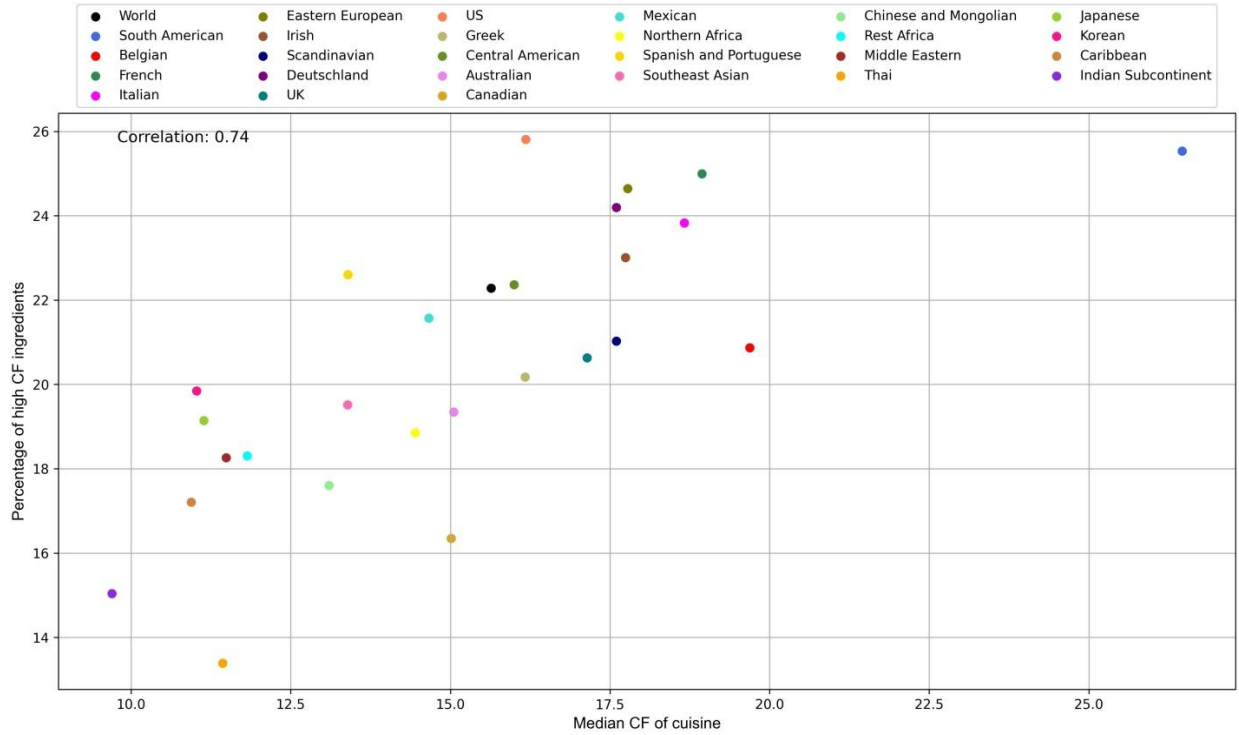

**Figure 2: Correlation of Top 1000 RecipeDB ingredients with High CF (CF>4) ingredients.** The median CF of cuisines shows a correlation of 0.74 for the Top 1000 most frequently used ingredients.

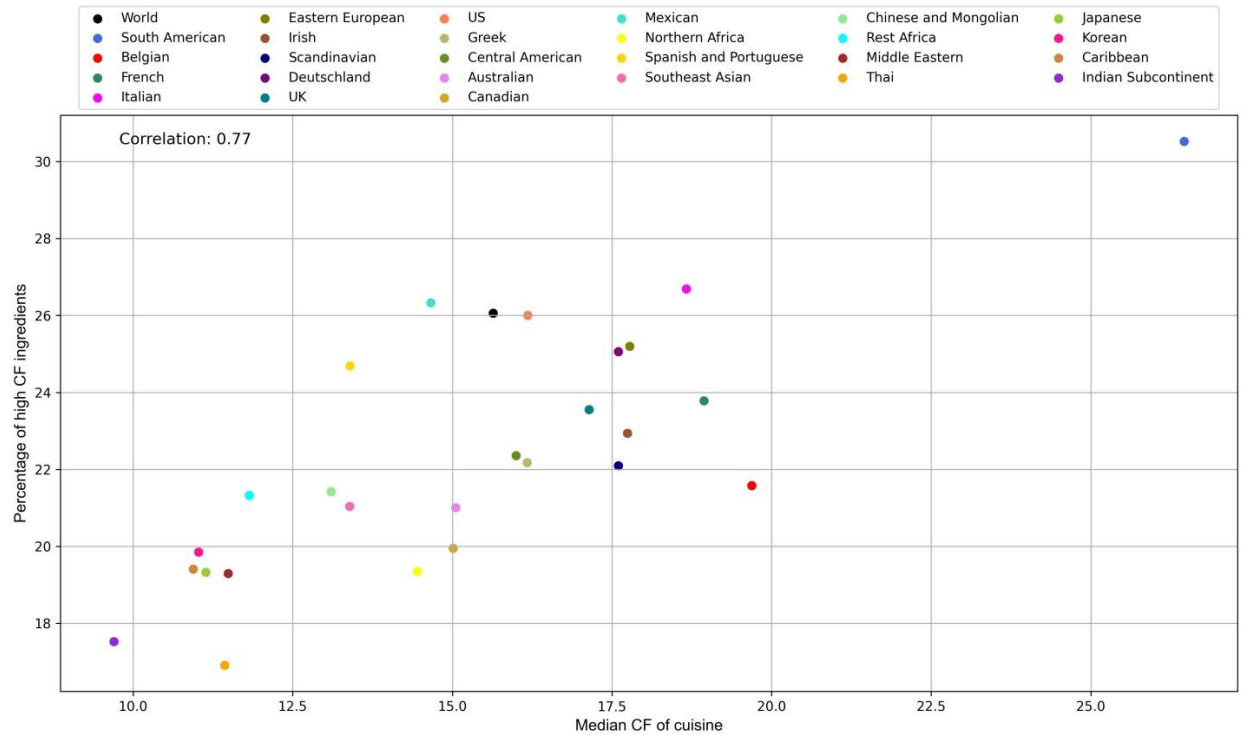

**Figure 3: Correlation of Top 10000 RecipeDB ingredients with High CF (CF>4) ingredients.** The median CF of cuisines shows a correlation of 0.77 for the Top 10000 most frequently used ingredients.

**List of unmapped SuEatable food products and food typologies**

The following SuEatable food products and food typologies were unmapped to any RecipeDB ingredient after BERT model implementation.

**Food Products**: cookies, crispbread, tempe, quorne, stracchino, asiago, camembert, emmental, pecorino, mealworms, mandarin, pomelo, tangerin, carob, rockmelon, maize, cowpea, alfonsino, anglerfish, flatfish, flathead, fork bread, ling, megrim, plaice, pomfret, porbeagle, rhombus, seabass, swordfish, turbot, whiting, barnacle, and krill.

**Food Typologies**: crispbread, legume, sweet, spices, shellfish, and tubers.

**Carbon load of a recipe**

$Carbon\ Load$ reflects the percentage of high CF ingredients (CF>4) in $a\ Recipe\ i$.

$$Carbon\ Load\ of\ a\ Recipe\ i = \frac{n_i^{CF>4}}{n_i^{mapped}}$$

$n_i^{CF>4}$ denotes the number of high CF ingredients in the recipe.
$n_i^{mapped}$ denotes the number of ingredients in the recipe successfully mapped to SuEatable.

Accordingly, averaging across all the recipes in a cuisine will yield the carbon load of the cuisine.